\documentstyle[preprint,aps,psfig]{revtex}
\begin{document}
\def\gev{{\rm \,Ge\kern-0.125em V}}  
\def\P{\Phi}
\def\bx{\bf {x}}
\def\half{{1\over2}}
\def\cM{{\cal M}}
\def\calT{\bar t}
\def\calB{{\cal B}}

\draft
\preprint{hep-ph/0301217}
\title{Out of Equilibrium Dynamics of the Inflaton Re-examined}
\author{Raghavan Rangarajan$^\dagger$ and Jitesh Bhatt$^*$
}
\address{Theoretical Physics Division, Physical Research Laboratory\\
Navrangpura, Ahmedabad 380 009, India}
\maketitle
\begin{abstract}
In a series of papers Boyanovsky et al. have studied the evolution of
an inflaton with a negative mass squared and a quartic self coupling
using the Closed Time
Path (CTP) formalism relevant for out-of-equilibrium dynamics.  
In this paper we comment on various aspects of these works.
We first compare their approach
to alternate approaches to study inflaton dynamics
and point out that the use of the CTP formalism gives the same
results as standard field theory in the Hartree and leading order
large N approximations.  We then rederive 
using the WKB approximation the large 
momentum mode functions of the inflaton needed for renormalisation
and point out some
differences with the previously obtained results.  
We also argue that the WKB approximation is valid only for large
$k/a$ and not for large $k$ as apparently assumed in the above mentioned
works.
We 
comment on the renormalisation prescription adopted in these
works and finally discuss how it differs from
another more commonly used prescription.

\vspace{0.7cm}
{\tt
\noindent $^\dagger$ raghavan@prl.ernet.in\\
$^*$ jeet@prl.ernet.in\\}
\end{abstract}
\newpage
\setcounter{page}{1} 
\section{Introduction}

In the new inflationary scenario the expansion of the universe is
driven
by the nearly constant potential energy of a slowly rolling scalar
field.  
In Refs. \cite{9610396,9709232}, Boyanovsky et al.
have discussed in great detail 
the evolution of 
an inflaton with a
negative mass squared and a quartic self coupling
using the Schwinger-Keldysh Closed Time Path (CTP)
formalism \cite{schwinger,keldysh} which is suited
for non-equilibrium or dynamical situations.
The advantage of the CTP formalism is that it allows one to
obtain $\langle in|in\rangle$ matrix elements of field operators
(as opposed to $\langle out|in\rangle$ matrix elements of standard
field theory) which are relevant in non-equilibrium or dynamical situations
\cite{cooperetal}.
\footnote{
Alternatively, in Ref. \cite{9310319} the authors use a Gaussian ansatz for the
density matrix, which is evolved in time with a time dependent Hamiltonian
using the von Neumann equation, and this is used
instead of the CTP formalism to obtain the
necessary expectation values for the dynamical system.}
In an earlier work \cite{bv} it had been shown (in the context
of Minkowski spacetime) that
if 
unstable modes grow large then a perturbative
treatment within the CTP formalism will be
insufficient and one needs to invoke a  non-
perturbative treatment.  Therefore in Ref. \cite{9610396,9709232} both the
Hartree and the large $N$ non-perturbative
approximations are used
to obtain the equation of motion for the inflaton.

Renormalisation was achieved in Refs. \cite{9610396,9709232}
by subtracting off the divergent ultraviolet
behaviour based on the renormalisation scheme employed in
Ref. \cite{9310319}.  
The ultraviolet behaviour
was obtained by analytically solving the equations of motion 
for
the large momentum fluctuations of the inflaton field 
after applying the Hartree or large $N$ approximations.
These approximations
render the equations of motion linear and thus amenable to 
techniques such as the WKB 
method.

In this work we first restate the equations of motion and the expression
for the inflaton fluctuations obtained in Refs. \cite{9610396,9709232}.
We then compare this
with earlier approaches to studying inflaton dynamics.
We next rederive the expressions for the large $k$ mode functions of the inflaton field 
needed for renormalisation,
after introducing
more
generalised initial conditions with finite temperature corrections
to the initial mass of the inflaton.
Since some of our results differ from those in Refs. \cite{9610396,9709232}
and related works, and since the calculations are involved, we have provided all
the details and 
pointed out
where we differ from previous works.
We have also commented upon certain aspects of
the renormalisation procedure. 
We then compare the renormalisation scheme adopted in Refs. \cite{9610396,9709232}
with schemes adopted by other authors.  Lastly we provide some numerical results.

To study the evolution of the inflaton in its potential
we choose initial conditions for the field modes consistent with
those specified in Ref. \cite{9709232}.
We work in the context of the large $N$ approximation and introduce an $N$-plet
$\vec \P$ 
to represent the inflaton.
As in Ref. \cite{9610396} we work in the context of a quenched
approximation 
in which the initial state of the universe at the onset of inflation
at $t=0$ corresponds to a state of positive mass $\P$ particles in 
thermodynamic equilibrium at a temperature $T_i>T_c$, the critical 
temperature,
while the subsequent
evolution is for a universe at $T=0$ in which the mass term for $\P$ 
becomes
negative.  We assume that the distribution of $\Phi$ particles freezes out
soon after the onset of inflation at a temperature close to $T_i$.

Our major conclusions are the following:
\begin{itemize}
\item 
The equations of motion obtained in
Refs. \cite{9610396,9709232}
afer invoking the CTP formalism 
are the same as those obtained using standard field
theory.
This is because of the
Hartree and lowest order
large N approximations adopted in 
Refs. \cite{9610396,9709232}.
\item The WKB solutions obtained in
Refs. \cite{9610396,9709232}
for large $k$ (i.e. $k\gg m_R$, where $m_R$ is 
the renormalised inflaton mass parameter)
are actually valid for large $k/a$ (i.e. $k/a\gg m_R$).  Since the 
renormalisation scheme of 
Refs. \cite{9610396,9709232}
involves subtracting the contribution of modes with $k>m_R$ the counterterms
used in
Refs. \cite{9610396,9709232}
do not match the bare quantities for $m_R a> k>m_R$.
\item The dominant modes relevant for inflaton evolution in new inflation
where the inflaton has a negative mass squared are the low momentum modes.
The counterterms used in
Refs. \cite{9610396,9709232} do not have any significant effect on the
contribution of these modes
and hence, despite our above criticism of the counterterms,
one gets the same results as in the more standard
prescription for removing ultraviolet divergences for the inflaton.
\item In the more standard prescription to deal with ultraviolet divergences
for the inflaton one subtracts the contribution of modes with $k/a>H$.
In Refs. \cite{9610396,9709232} the intent of the renormalisation scheme appears
to be to subtract the contribution of modes with $k>m_R$.  While for the case
of an inflaton with a negative mass squared it turns out that both schemes give
the same results, for an arbitrary field the renormalisation scheme of
Refs. \cite{9610396,9709232} should be used with care.  We illustrate this with
an example of a massless field in de Sitter space.
\end{itemize}

\section{Equations of motion and initial conditions for inflaton
modes, and total fluctuations}

The 
Lagrangian density that we consider in the context of 
a spatially flat Robertson-Walker universe is given by,
\begin{equation}
{\cal L}  =   a^3(t)\left[\frac{1}{2}\dot{\vec{\Phi}}^2(x)
-\frac{1}{2}
\frac{(\vec{\nabla}\vec{\Phi}(x))^2}{a^2(t)}
-V(\vec{\Phi}(x))\right],
\label{lagrangian} 
\end{equation}
\begin{equation}
V(\vec{\Phi})  =  
{Nm^4\over2\lambda} +\frac{1}{2}M_{g}^2\vec{\Phi}^2 
+ \frac{\lambda}{8N} \vec{\Phi}^4 
\end{equation}
$\vec{\Phi}$ is an $N$-plet
and $M_{g}^2  = - m^2+\xi\;{\cal R}$.
$\cal R$ is the Ricci scalar and is given by
\begin{equation}
{\cal R}  =  6\left(\frac{\ddot{a}(t)}{a(t)}+
\frac{\dot{a}^2(t)}{a^2(t)}\right). \label{ricciscalar}
\end{equation}
The scale factor can be obtained dynamically using the Einstein equation.
(In the numerical solutions
in Ref. \cite{9709232} $H$ was obtained dynamically
while it was taken as a constant in the simulations of Ref. \cite{9610396}.)

Below we sketch 
the large $N$ approximation as discussed in
greater detail in
Refs. \cite{9408214,9610396}.  We then present the 
equations of motion and discuss the initial conditions.  
In this section we also discuss other approaches to study inflaton dynamics.

Non-perturbative treatment of non-equilibrium dynamics
can be studied by approximations such as the Hartree approximation
and the large $N$ expansion \cite{9610396}. 
These approximations are similar at leading order in $1/N$.
Here we shall  work with the large $N$ approximation.
In the large $N$ approximation the Lagrangian
can be written in terms of
$\lambda/N$ as above
and in the spirit of Ref. \cite{9610396} 
it is assumed below that $\lambda$ is small
enough that 
the large $N$ approximation is valid even when
$N=O(1)$.  (We shall choose $\lambda=10^{-12}$ later.)
Following Ref. \cite{9408214,9610396}, only 
leading order terms in the large $N$ approximation are retained below.

To facilitate the use of the large $N$ approximation to study the
inflationary phase transtition the inflaton 
is written as
$$
\vec{\Phi}(\vec x, t) = (\sigma(\vec x,t), \vec{\pi}(\vec x,t)),
$$ 
where $\vec{\pi}$ is an $N-1$-plet, and we let
\begin{equation}
\sigma(\vec x,t) = \sigma_0(\vec x,t) 
 + \rho(\vec x,t) \; \; ; \; \; 
\langle
\sigma(\vec x, t) \rangle= \sigma_0(\vec x,t) \; \; ; \; \; \langle 
\rho(\vec x,
t) \rangle = 0.
\label{largenzeromode} 
\end{equation}
To apply the large $N$ approximation $\sigma_0$ and $\vec{\pi}$ are defined by
\begin{equation} 
\sigma_0(\vec x, t)=\sqrt{N}\phi(t) \; , \;
\vec{\pi}(\vec x, t)=\psi(\vec x, t)
\overbrace{\left(1,1,\cdots,1\right)}^{N-1} \;, \;
\end{equation} 
$\phi$ mimics the mean of the inflaton field
while $\psi$ 
and $\rho$ 
correspond to fluctuations.  
It is implicitly assumed here that the mean field is a function of time only.

The mode functions $U_k(t)$ of the inflaton are defined as
\begin{eqnarray}
\psi(x)&=&\int \frac{d^3k}{(2\pi)^3 \sqrt 2} [U_k(t) 
a_{\bf k}e^{i{\bf k}.{\bf x}}
+ h.c.]
\end{eqnarray}
with $[a_{\bf k},a^\dagger_{{\bf k}'}]
=(2\pi)^3
\delta({\bf k}-{\bf k}')$.
The leading order in $1/N$
equations of motion for $\phi$ and 
$U_k(t)$ 
then are
\begin{equation}
\ddot{\phi}(t)+3H\dot{\phi}(t)+M_{g}^2\phi(t)+\frac{\lambda}{2}\phi^3(t)+
\frac{\lambda}{2}
\phi(t)\langle \psi^2(t)\rangle
=0, \label{largezeromodeeqn}
\end{equation}
\begin{equation}
\left[\frac{d^2}{dt^2}+3H\frac{d}{dt}+\omega^2_k(t) \right]U_k(t)= 0,
\label{largenmodes}
\end{equation}
and the effective frequency $\omega_k(t)$ is given by
\begin{equation}
\omega^2_k(t) =\frac{k^2}{a^2(t)}+M^2(t) \; ,
\label{largenfreq}
\end{equation}
where
\begin{equation}
M^2(t) =  M_{g}^2+ \frac{\lambda}{2}\phi^2(t)+
\frac{\lambda}{2}\langle \psi^2(t) \rangle 
\; ,
\label{Ngranmass}
\end{equation}
and
\begin{equation}
\langle \psi^2(t) \rangle = \int
\frac{d^3k}{(2\pi)^3}\frac{|U_k(t)|^2}{2}{\rm coth}[ W_{k0}/(2T_i)].
\label{psisqunren}
\end{equation}
By  $\langle \psi^2(t) \rangle$ 
we mean $\langle \psi^2({\vec x},t) \rangle(t)$. 
By translational invariance
$\langle \psi^2({\vec x},t) \rangle(t)$ depends only on time.
 
The coth function reflects the initial thermal particle distribution
and it is 
assumed that 
$\psi$ goes 
out of equilibrium at a 
temperature close to $T_i$.  
We obtain $T_i$ by specifying that inflation starts when
the energy density in radiation falls to one-tenth of the 
constant term in the 
potential energy.  Assuming the number of relativistic degrees
of freedom to be 100, we get $T_i\sim 200 m_R$.
$ W_{k0}$
is related to the initial frequencies of the modes and is defined later.

At leading order in $1/N$,
$\rho$ 
does not enter in the equations of motion for 
$\phi$ and
$U_k$. 
Furthermore the contribution of 
$\rho$ 
to the energy density is smaller by a power of 
$N$ compared to that of the other fields.  Hence 
it is not included in the analysis.

The above equations of motion are similar to those in earlier works that do not
invoke the CTP formalism, such as Ref. \cite{vilenkin}.  
Modifying Eqs. (2.13) and (2.14) of Ref. \cite{vilenkin} for our potential 
(and our definition of the Fourier modes)
gives
\begin{equation}
\ddot U_k +3H \dot U_k +\biggl(\frac{k^2}{a^2} +M_g^2 
+3\frac{\lambda}{2}\langle \psi^2(t) \rangle^\prime \biggr) U_k=0
\label{vil}
\end{equation}
where
\begin{equation}
\langle \psi^2(t) \rangle^\prime \equiv 
{_{ out}\langle} \psi^2(\vec x,t) \rangle_{in}=
 \int
\frac{d^3k}{(2\pi)^3}\frac{|U_k(t)|^2}{2}.
\end{equation}
The similarity in our equations and those 
of
Ref. \cite{vilenkin}
\footnote{
The factor of 3 with $\lambda$ above is because the cactus approximation in
Ref. \cite{vilenkin} is analogous to the Hartree approximation, and the 
leading order large
$N$ and Hartree approximations differ by this factor of 3 \cite{9610396}.
$\phi(t)$ is absent in the equation of motion for $U_k$
as the mean field is absent in Ref. \cite{vilenkin}, and
temperature effects are ignored.},
though the latter are obtained without invoking the CTP formalism, is 
because 
we are working to lowest order in $1/N$, and at this order
both the $\langle out|in\rangle$ and the $\langle in|in\rangle$
formalisms give the same equations of motion. 
This may be verified by looking at the field equations
obtained using standard field theory and the CTP formalism
in Ref. \cite{cooperetal}.  Eqs. (4.27)-(4.34) of Ref. \cite{cooperetal}
give the equations of motion in the CTP formalism including terms of
order $1/N$
for the mean field and the momentum modes and these
reduce to equations obtained using standard field theory, namely,
Eqs. (2.20) and (2.21) of Ref. \cite{cooperetal} for the mean field
and Eq. (\ref{vil}) above for the momentum modes, if we ignore terms of order $1/N$.
\footnote{
Curved spacetime terms are absent in 
Ref. \cite{cooperetal}.
}
Note that at lowest order in $1/N$ 
the equal time Wightman functions 
$i\langle \psi^2(t)\rangle$ obtained using the CTP formalism
(also referred to as $G_>$,$G_<$)
and the equal time
Feynman propogator obtained in standard field theory
that appear in the equations of motion
are the same (see Sec. II of Ref. \cite{9610345}).

Another approach to studying inflaton dynamics is provided in 
Ref. \cite{hawkingmoss2} in which the mean field $\langle\P\rangle$ is 
taken
to be 0 and the dynamics is studied by evolving $\langle\P^2\rangle$
by defining an effective action which is a function of
$\langle\P^2\rangle$ rather than of $\langle\P\rangle$, 
similar to the Cornwall-Jackiw-Tomboulis method \cite{CJT}
in which one studies the two particle irreducible effective action.  
In Ref. \cite{hawkingmoss2} the authors do not invoke the CTP
formalism.  In Ref. \cite{calzetta} the CJT method is combined
with the CTP formalism to study the evolution of a $\lambda \phi^4$
scalar field with a tachyonic mass which is suddenly brought into contact
with a heat bath at zero temperature (but in Minkowski spacetime).
The CJT method directly gives an equation of motion for the two-point
function.  In contrast, above, as in 
Refs. \cite{9610396,9709232,cooperetal}, we obtain
the time-dependent two-point function by solving for the mode functions
$U_k$ and then substitute them in the expression for the two-point 
function
Eq. (\ref{psisqunren}).
Note also when comparing the above approaches
that when one uses the large $N$
approximation one is expanding in powers of $\lambda/N$ while in the 
CJT
method one is expanding in powers of $\hbar$.  The reader is also 
referred to Ref. \cite{dasbook2} in which the time-dependent two-point 
function
is obtained for a system undergoing a phase transition with a quench 
albeit
in Minkowski space time and without a time-dependent effective mass
after the quench.

{\bf Initial conditions:}

As discussed in
Ref. \cite{9709232}, 
for $\phi(0)\ll  H$
and $\dot\phi(0)\approx0$ 
the growth of $\phi$ does not contribute much to 
the evolution of the inflaton.
We assume $\phi(0)$=0 and $\dot\phi(0)=0$
which implies that the mean field
does not evolve with time.  This is consistent with the argument presented in
Ref. \cite{hawkingmoss2} that for a symmetric potential in de Sitter
space the expectation values of the fields will be 0 in the absence
of sources because the finiteness of the effective spatial volume due 
to the presence of an event horizon allows for
tunneling between the states with expectation value
$\phi$ and $-\phi$. 
\footnote{
Timescales for this tunneling and thermal initial states are not
discussed in Ref. \cite{hawkingmoss2}.
The non-zero
overlap between the vacua associated with the 
two minima of our potential when the volume is finite 
may also be deduced from the discussion in
Ref. \cite{umezawaetal}.  It
is because the vacua are related by
a non-zero transformation when the volume is finite.  
}
Note that 
if $\phi(t)=0$ 
then the field $\Phi$ can
be represented as an $N$-plet of the form $\psi^{''}\overbrace{(1,1,...)}^N$.  
However, in the
large $N$ limit, working with an $N-1$-plet as we do gives the same
results.

We assume that the initial size 
of the inflating 
region
is $O(H^{-1})$ and 
only include
modes with $k/a_0\ge 2H(0)$ in our expressions 
for $\langle\psi^2\rangle$,
energy $\epsilon$, pressure $p$, etc. \cite{lindebook}. 
(In Refs. \cite{9610396,9709232} all modes are included.  Our numerical 
results indicate that there is no significant effect of our lower momentum
cutoff.)
Then, as discussed further below, 
the initial conditions for $U_k$ for modes with $k/a_0\ge 2 H(0)$
are given
by
\begin{eqnarray}
U_k(0) &=& \frac{1}{\sqrt {W_{k0}}} \; \; ; \; \dot{U}_k(0)=
\left[-iW_{k0}-H(0)\right] U_k(0), \label{inicondcomo1}
\end{eqnarray}
where
\begin{eqnarray}
W_{k0}^2=k^2+
a_0^2[(-1+r^2)m^2 +(\xi-{1\over6}) {\cal R}(0) + 
\frac{\lambda}{2}\phi^2(0)+
\frac{\lambda}{2}\langle \psi^2(0) \rangle 
]\; .
\end{eqnarray}
The term $r^2 m^2$ represents the 
corrections to the initial inflaton mass, predominantly thermal corrections,
due to the interactions of the inflaton with other fields.
We assume $
\frac{\lambda}{2}\langle \psi^2(0) \rangle$ is small
compared to other terms in $W_{k0}^2$ and so ignore
it while entering the initial conditions in our numerical
programme.
Our choice of the factor 2 in the lower momentum cutoff
ensures that the initial frequencies
defined above are real with our choice later of $H(0)=2m$
for all relevant values of $k$ 
irrespective
of the value of $r$, and that the
contributions related to the initial fluctuations in $\psi$
can indeed be ignored in the initial
frequencies.  (The existence of the low momentum cutoff
allows us to avoid the issue of imaginary frequencies at $t=0$ for
lower momentum modes, which has to be dealt with in Ref. \cite{9709232}.)

Refs. \cite{9610396,9709232} differ
in the initial conditions for the mode functions, as discussed in 
the section on the WKB approximation.
Refs. \cite{9610396} and \cite{9709232} also differ in the initial 
temperature
taken in the initial conditions for
the inflaton.  In Ref. \cite{9610396}, the authors studied a quenched 
approximation from an initial temperature $T_i>T_c$ to $T\sim 0$.
The initial conditions for the modes of the inflaton field are 
specified at temperature
$T_i\sim10^7 m_R$.  In Ref. \cite{9709232}, the initial 
temperature is taken to be 
0.  As
was noted in Sec. IVA of Ref. \cite{9610396}, this gives a much longer 
duration
of inflation.  
$T_i$ enters in both the coth function and the choice of $r$.  Below
we show that it is the influence via the former that causes a change in
the duration of inflation.  This may also be deduced from the arguments of 
Ref. \cite{9610396}.

\section{The large ${k}$ solution and renormalisation}

{\bf WKB:}

We now discuss the WKB approximation used
to obtain the mode functions and their derivatives for large values of 
$k$.
The renormalised $\langle\psi^2\rangle$,
$\epsilon$ and $\epsilon + p$ are then obtained by subtracting the contribution
of the large momentum modes.
Our calculation is similar to that of Ref. \cite{9703327}.
We provide a complete derivation so as to
include some details not provided in the literature.  Furthermore
our results differ from those in the literature and we point out
the differences.

The large $k$ mode functions for a scalar field $\Phi$ with a quartic
potential 
were obtained in Ref. \cite{bv} for evolution in Minkowski space and 
later in
Refs. \cite{9310319} and \cite{9610396} for evolution in
FRW and de Sitter spacetimes respectively.
These were reobtained in Refs. 
\cite{9703327,9709232}
with a different set of initial conditions than 
in Refs. \cite{9310319,9610396}.  The new initial
conditions for the mode functions were obtained by first defining them
for the field modes in conformal spacetime.
With the new initial conditions,
Ref. \cite{9703327} focussed on radiation
and matter dominated backgrounds while
the inflationary era was studied in Ref. \cite{9709232}.
Below we obtain the large $k$ mode functions for 
$\Phi$
with initial conditions similar to those in 
Refs. \cite{9709232}
and further generalise the initial conditions to
include high temperature corrections to the initial effective mass,
i.e., the $r^2 m^2$ term in $W_{k0}$.
The latter
was included in Ref. \cite{9610396} but not in Ref. \cite{9709232}.
In Refs. \cite{9703327,9709232} it was stated that the large $k$ behaviour of 
the mode functions, which is used to define the counterterms for
renormalisation, is independent of initial conditions if one
chooses the new initial conditions. 
As we see below, this is true only 
if temperature corrections to the inflaton mass at the
onset of inflation are 0.

To obtain the large $k$ mode functions, we work in conformal
spacetime. 
Defining the conformal time $\tau$ from $d\tau=dt/a(t)$ 
the mode functions in conformal time are defined via
$U_k(t)=f_k(\tau)/C(\tau)$
where $C(\tau)=a(t)$ when $\tau$ and 
$t$ are related as above.
We have chosen the normalization of the scale factor such that
$a(0)= C(\tau_0)=1$. 
 
The equation of motion for $f_k$ is 
(see the Appendix of Ref. \cite{9703327} for
more details)
\begin{eqnarray}
\left[\frac{d^2}{d\tau^2}+k^2+{\cal M(\tau)}^2 \right]f_k(\tau)&=& 0
\label{fkeom}
\end{eqnarray}
with
\begin{eqnarray}
{\cal M}^2=C^2[-m^2 +(\xi-{1\over6}) {\cal R}] 
+ \frac{\lambda}{2}\tilde\phi^2+
\frac{\lambda}{2}\langle \tilde\psi^2 \rangle 
\label{Msqconf}
\end{eqnarray}
where the tilde refers to the conformally rescaled field, i.e.,
$\phi(x,t)=\tilde\phi(x,\tau)/C(\tau)$, etc.  
$R(\tau)=6C^{\prime\prime}(\tau)
/C^3(\tau)$ where $'$ refers to differentiation with respect to $\tau$.  
The above expression for the effective mass
differs from that in Eq. (A7) in Ref. \cite{9703327}.

The initial conditions for the conformal time
mode functions $f_k$ 
are chosen to 
correspond to an initial density matrix that represents a system in
local thermodynamic equilibrium and which commutes with the Hamiltonian in
conformal spacetime
at $\tau=\tau_0$ \cite{9703327}.  Therefore, for modes with $k/a_0\ge 2H(0)$,
\begin{equation}
f_k(\tau_0)= \frac{1}{\sqrt{{\cal W}_{k0}}} \; ; \; \dot{f}_k(\tau_0) = 
-i {{\cal W}_{k0}} f_k(\tau_0)
\; ; \;
{\cal W}_{k0}= \sqrt{k^2+{\cal M}^2_0}. \label{initcondpsi}
\end{equation}   
${\cal M}_0$ is the initial effective mass
in terms of conformal time.
\begin{eqnarray}
{\cal M}^2_0=C^2(\tau_0)[(-1+r^2)m^2 +(\xi-{1\over6}) {\cal R}(\tau_0)] + 
\frac{\lambda}{2}\tilde\phi^2(\tau_0)+
\frac{\lambda}{2}\langle \tilde\psi^2(\tau_0) \rangle 
\end{eqnarray}
Note that the temperature corrections
$r^2m^2$ 
above is not included in Eq. (\ref{Msqconf})
which is valid at times after the quench.  
The presumption here is that the other fields to which the inflaton
couples go out of equilibrium during the quench at a temperature much
below $T_i$ at which stage their contribution to the effective
mass of the inflaton is small and can be ignored, and that any subsequent
growth in their fluctuations during inflation
does not lead to a significant contribution to the effective
inflaton mass.

These initial conditions for modes with $k/a_0\ge 2H(0)$
translate to 
\begin{eqnarray}
U_k(0) &=& \frac{1}{\sqrt {W_{k0}}} \; \; ; \; \dot{U}_k(0)=
\left[-iW_{k0}-H(0)\right] U_k(0), \label{inicondcomo}
\end{eqnarray}
where $W_{k0}$ 
is ${\cal W}_{k0}$ 
expressed in comoving 
coordinates.

Dividing the equation of motion for $f_k$ Eq. (\ref{fkeom}) 
by $k^2/m^2$, we treat $m/k\equiv\delta$ as a small
parameter and apply WKB theory to solve for $f_k$.  
We shall seek solutions for $k\ge 10m$.  
Using a WKB solution for $f_k$ of the form \cite{benderorszag} 
\begin{equation}
D_k(\tau) \sim {\rm exp}[{1\over\delta}S(\tau)] 
={\rm exp}[{1\over\delta}\sum_{n=0}^\infty \delta^n S_n]
\equiv e^{\int^{\tau}_{\tau_0} R_k(\tau')d\tau'}\, ,
\label{Dk}
\end{equation}
we obtain two solutions, $D_{k1}(\tau)$ and its complex conjugate,
which agree with the expressions in Eqs. (A14) and (A15) of Ref. 
\cite{9703327}, namely,
\begin{equation}
R_k(\tau) = -ik+R_{0,k}(\tau)-i\frac{R_{1,k}(\tau)}{k}+
\frac{R_{2,k}(\tau)}{k^2}-i\frac{R_{3,k}(\tau)}{k^3}+
\frac{R_{4,k}(\tau)}{k^4}+ \cdots
\label{Rk}
\end{equation} 
and its complex conjugate.  The coefficients are given by
\begin{eqnarray}
&& R_{0,k} = 0 \; \; ; \; \; R_{1,k} = \frac{1}{2} {\cal{M}}^2(\tau) \; 
\; ; \;
\; R_{2,k} = -\frac{1}{2} R'_{1,k}=-\frac{1}{4} 
{\cal{M}}^{2\prime}(\tau) 
\nonumber \\ &&R_{3,k} = \frac{1}{2}\left(
R'_{2,k}-R^2_{1,k} \right)=
-\frac{1}{8} {\cal{M}}^{2\prime\prime}(\tau)-\frac{1}{8} 
{\cal{M}}^{4}(\tau)\nonumber\\
&&R_{4,k} = -\frac{1}{2}\left(
R'_{3,k}+2R_{1,k}R_{2,k} \right) 
=\frac{1}{16} {\cal{M}}^{2\prime\prime\prime}(\tau)+\frac{1}{8} 
{\cal{M}}^{4\prime}(\tau)\,.
\end{eqnarray}      
Imposing the initial condition for $f_k(\tau_0)$ implies
\begin{equation}
f_k(\tau) = \frac{1}{2\sqrt{{\cal W}_{k0}}}\left[
   (1+\gamma)D_{k1}(\tau)+(1-\gamma)D^*_{k1}(\tau) \right] \,.       
\end{equation}
$\gamma$ is then obtained by imposing the initial condition on
$f_k^\prime(\tau_0)$.  We find that the real and imaginary parts of 
$\gamma$ are given by
\begin{eqnarray}
\gamma_R&=& 1+{g_1\over k^2} + {g_2\over k^4}\cr 
\gamma_I&=& {g_3\over k^3}+{g_4\over k^5}
\end{eqnarray}
where
\begin{eqnarray}
g_1&=&{1\over2}({\cal M}^2_0-{\cal M}^2_{0+})\cr
g_2&=&{1\over8}{\cal M}^{2\prime\prime}_{0+}-{1\over8}({\cal 
M}^4_0-{\cal M}^4_{0+})
-{1\over4}{\cal M}^2_{0+}({\cal M}^2_0-{\cal M}^2_{0+})\cr
g_3&=&{\cal M}^{2\prime}_{0+}\cr
g_4&=&-{1\over16}{\cal M}^{2\prime\prime\prime}_{0+}
-{1\over2} {\cal M}^2_{0+} {\cal M}^{2\prime}_{0+}+{1\over8}{\cal 
M}^2_0{\cal M}^{2\prime}_{0+}
\end{eqnarray}
and ${\cal M}^2_{0+}\equiv{\cal M}^2(\tau_0)$.  Note that
${\cal M}^2_{0}-{\cal M}^2_{0+}=r^2m^2$ since
${\cal M}^2_{0}$ and ${\cal M}^2_{0+}$ 
are the effective masses in conformal
spacetime just before and after the onset of inflation.

Expressing $D_{k1}(\tau)$ as $e^{X+iY}$ and using the above relations
we get
\begin{equation}
|f_k|^2=\frac{1}{{\cal W}_{k0}}e^{2X}[\cos^2Y+\gamma_R^2\sin^2Y
-2\gamma_I\cos Y\sin Y +\gamma_I^2 \sin^2Y]
\label{fkmodsq1}
\end{equation}
Ultimately the above expression will be integrated over all $\bf k$.
Expressing trignometric functions as exponentials there are terms
in the integrand that for large $k$ go as 
$k^2 k^{-1-l} {\rm exp}[\pm 2ik(\tau-\tau_0)$, $l\ge0$.  For large
$k(\tau-\tau_0)$ we set their integrals to 0.
So
we replace 
$\cos^2Y$ and
$\sin^2 Y$ by $1\over2$ and $\cos Y\sin Y$ by 0  
and write the effective $|f_k|^2$ as
\begin{eqnarray}
|f_k|^2&=&\frac{1}{2{\cal W}_{k0}}e^{2X}[1+\gamma_R^2]\cr
&=&\frac{1}{k}\biggl[1-{1\over k^2}{{\cal M}^2(\tau)\over2}
+{1\over k^4}\biggl\{ {\cM^{2\prime\prime}(\tau)\over8} +
{3\cM^{4}(\tau)\over8} + {(\cM^2_0-\cM^2_{0+})^2\over8} \biggr\}\biggr]
\, ,\label{fkmodsq}
\end{eqnarray}
where we have also ignored the term proportional
to $\gamma_I^2$ since it is of order $1/k^6$ and can be ignored in the 
large $k$
limit.
$|f_k^\prime|^2$ is given by
\begin{eqnarray}
|f_k^\prime|^2=\frac{1}{{\cal W}_{k0}}e^{2X}\biggl[
\bigl \{
R_{kR} &\cos Y -R_{kI} \sin Y -\gamma_I(R_{kR} \sin Y +R_{kI} \cos 
Y) 
\bigr\}^2 \cr
&+ \gamma_R^2\bigl\{R_{kR} \sin Y + R_{kI} \cos Y \bigr\}^2
\biggr ]\, ,
\label{fkprimemodsq}
\end{eqnarray}
where $R_{kR}$ and $R_{kI}$ are the real and imaginary parts of $R_k$
defined in Eq. (\ref{Dk}).
Again, the effective $|f_k^\prime|^2$ relevant for subsequent 
integration over
all $\bf k$ and excluding terms of higher order than $1/k^3$ is
\begin{eqnarray}
|f_k^\prime|^2&=&\frac{1}{2{\cal W}_{k0}}e^{2X}[1+\gamma_R^2]R_{kI}^2\cr  
&=&k\biggl[1+{1\over k^2}{{\cal M}^2(\tau)\over2}\cr
&&+{1\over k^4}\biggl\{-{\cM^{2\prime\prime}(\tau)\over8}
-{\cM^{4}(\tau)\over8} + {(\cM^2_0-\cM^2_{0+})^2\over8}
+{\cM^{2}(\tau)\over4}(\cM^2_0-\cM^2_{0+})\biggr\}
\biggr]
\end{eqnarray}
(We obtain $|f_k|^2$ to $O(1/k^5)$ and $|f_k^\prime|^2$ to $O(1/k^3)$ 
since
the former is multiplied by a factor of $k^2$ in the expression for 
energy
and pressure. However in the expression for $\langle\psi^2\rangle$ we 
only
keep terms to $O(1/k^3)$ in $|f_k|^2$.)

We reiterate that the terms in Eqs. (\ref{fkmodsq1}) and
(\ref{fkprimemodsq}) set to 1/2 and 0
attain these values only after integration over all $\bf k$.
There is no mention of this in the literature.  
Furthermore, during inflation $\tau=-1/(aH)$.  Treating $H$ as a 
constant equal to $2m$ we can easily check that for
$k\ge3m$, $k(\tau-\tau_0)>>1$ is valid within two
e-foldings or so of inflation.  
Once inflation ends the universe
expands as $t^n$ ($n<1$) and $\tau=\frac{n}{1-n}(aH)^{-1}$.  Then  
$\tau$
is positive and increases with time.  Hence $k(\tau-\tau_0)>>1$
will continue to be valid.
However for $k\le2m$,
$k(\tau-\tau_0)>>1$
will never hold during the inflationary era.
Therefore later we shall use the above large $k$ expressions only for
$k\ge10m_R$, where $m_R$ is the renormalised mass parameter, 
rather than for $k\ge m_R$ as in Refs. \cite{9610396,9709232}, while
subtracting off the divergences during renormalisation.
 
As mentioned earlier, one can see that it is only for $r=0$,
i.e. when $\cM^2_0=\cM^2_{0+}$, 
that the large $k$ subtractions
needed to renormalise the 
divergences are independent of initial conditions.
For $r=0$, our result for $|f_k(\tau)|^2$ agrees with
Eq. (A24) of Ref. \cite{9703327}.  Our result for
$|f_k^\prime(\tau)|^2$ disagrees with Eq. (A25) of Ref. 
\cite{9703327}.  However it agrees with the corresponding 
expression in Eq. (5.24) of the review article 
\cite{0006446} indicating a possible typographical 
error in Ref. \cite{9703327}.  (Note that the expression for
$|f_k(\tau)|^2$ ($r=0$) in Eq. (5.24) of  
Ref. \cite{0006446} disagrees with both our expression 
and that of Ref. \cite{9703327} indicating a possible 
typographical error in Ref. \cite{0006446}.)

The corresponding expressions for the comoving time mode
functions are
\begin{eqnarray}
|U_k(t)|^2 & = & \frac{|f_k(\tau)|^2}{C^2(\tau)} \\
&=& \frac{1}{ka^2(t)}- \frac{1}{2k^3
a^2(t)}\;B(t)  \cr &&+
{1 \over {8 k^5 \; a^2(t) }}\biggl[  3 B(t)^2 + a(t)\dot a(t)\dot B(t)
+a(t)^2 \ddot B(t)+(B_0-B_{0+})^2
  \biggr] +
{\cal{O}}(1/k^7)  \cr
& = & {\cal S}^{(2)}+ {\cal{O}}(1/k^7) \; ,
\label{Ukmodsq}\\                              
|\dot{U}_k(t)|^2 & = & \frac{1}{C^2(\tau)}
\left[ \frac{|f'_k(\tau)|^2}{C^2(\tau)}+ \left(H^2-\frac{H}{C(\tau)}
\frac{d}{d\tau}\right)|f_k(\tau)|^2 \right]\\
&=&
\frac{k}{a^4(t)}+\frac{1}{2ka^4(t)}\left[B(t)+2\dot{a}(t)^2 \right] 
\cr
&  &+ {1 \over {8 k^3 \; a^4(t) }}\biggl[ - B(t)^2 - a(t)^2 {\ddot
B}(t) + 3 a(t)
{\dot a}(t)
{\dot B}(t) - 4 {\dot a}^2(t) B(t) + (B_0-B_{0+})^2 
\cr&&
+2B(t)(B_0-B_{0+})  \biggr] +  {\cal{O}}(1/k^5) \cr
& = & {\cal S}^{(1)}+ {\cal{O}}(1/k^5) \; ,
\label{Ukdotmodsq}
\end{eqnarray}    
where $B(t)$ is ${\cal M}^2(\tau)$ expressed in comoving coordinates
and $B_0$ and $B_{0+}$ are similarly
defined.

{\bf Renormalisation:}

As in Ref. \cite{9610396,9709232} we use
the above expressions to write the renormalised
$\langle\psi^2\rangle$, $\langle\psi^2\rangle_R$, as
(hereafter the subscript $R$ refers to renormalised quantities)
\begin{equation}
\langle \psi^2(t) \rangle_R = \int
\frac{d^3k}{2(2\pi)^3}
{\rm coth}[{ W_{k0}}/(2T_i)]
\left[
|U_k(t)|^2
\Theta(k-2H(0))-\Theta(k-\kappa)\left(
\frac{1}{ka^2(t)}- \frac
{B}{2k^3a^2}
\right)
\right] \, .
\label{psisqR}
\end{equation}
$\kappa$ represents a renormalisation scale.  As mentioned above, 
the large $k$ expressions used in the subtraction are valid for
$k\ge10m_R$ (after two e-foldings).  Hence we 
take $\kappa$
to be $10 m_R$ rather than $m_R$ as in Ref. \cite{9709232}.  
We also choose to not rescale $\langle\psi^2\rangle_R$ to be 0 at 
$t=0$.
As in Ref. 
\cite{9610396}
we shall take $H(0)=2 m_R$.  

$B(t)$ is now expressed in terms of renormalised parameters
making use of the condition, $M^2_R(t)=M^2(t)$
discussed
in Sec. IV of Ref. \cite{9310319}.  (Note that $M^2(t)$ 
is written in terms of bare
parameters.)
Therefore
\begin{equation}
B(t)=a^2(t)
[-m_R^2 +(\xi_R-{1\over6}) {\cal R}(t) + 
\frac{\lambda_R}{2}\phi^2(t)+
\frac{\lambda}{2}\langle \psi^2(t) \rangle_R 
]  
\end{equation}
Since $B$ contains $\langle \psi^2 \rangle_R$ one has to solve
Eq. (\ref{psisqR}) to obtain an explicit expression for 
$\langle \psi^2\rangle_R$.  

Besides the presence of the first $\Theta$ function,
our approach above also differs from that of Ref. \cite{9610396,9709232}
in that the second $\Theta$ function applies to all terms in the 
subtraction
and not just to the last term that suffers from an infrared singularity.  
\footnote{
The expression for $\langle \psi^2(t) \rangle_R$ in Eq. (3.18)
of Ref. \cite{9610396} is missing
one subtraction term (and the coupling).
We thank D. Cormier for confirming that this is a
typographical error.  There are similar errors in Eq. (3.20)
of Ref. \cite{9610396} and in the expression for the renormalised
fluctuations in Sec. V of Ref. \cite{9709232}.
} 
The
difference between the subtractions under the prescription above
and that of Refs. \cite{9610396,9709232} is finite and so both are 
valid.
However, 
as we discuss below, 
a prescription that subtracts all terms in an expansion for $|U_k|^2$
for large $k$ modes
makes it simpler to obtain
the expressions for the renormalised energy and energy plus 
pressure
that satisfy the energy conservation equation.

Similar to the approach in Refs. \cite{9610396,9709232},
the renormalised energy and energy plus pressure are defined as
\footnote{
There is an error in the last term in the expression for the 
renormalised energy density in Eq. (3.21) of Ref. \cite{9610396}.}
\footnote{
An alternate prescription of obtaining the renormalised energy-momentum
tensor using adiabatic subtraction is provided in 
Ref. \cite{finellietal}, albeit for chaotic inflation.  In this prescription
one expands bare quantities in powers of the derivatives of ln$a$ which
is equivalent to expanding about solutions for Minkowski spacetime.}
\begin{eqnarray}
{\epsilon_R\over N}&=&
{m_R^4\over 2\lambda}+
\frac{1}{2}\dot\phi^2 + 
\frac{1}{2}M_{gR}^2\phi^2 + 
\frac{\lambda_R}{8}\phi^4\nonumber\\
&&+\frac{1}{2}\int \frac{d^3k}{2(2\pi)^3}{\rm coth}[{ W_{k0}}/(2T_i)]
\left[|\dot U_k|^2\Theta(k-2H(0))-{\cal S}^{(1)}\Theta(k-\kappa) 
\right.\nonumber\\&&
\left.+\omega_{kR}^2 (|U_k|^2\Theta(k-2H(0))
-{\cal S}^{(2)}\Theta(k-\kappa)  )
\right]-
\frac{\lambda_R}{8} \langle\psi^2\rangle^2_R\nonumber\\
\end{eqnarray}
\begin{eqnarray}
{(p+\varepsilon)_R\over N}& = & 
\dot{\phi}^2 
+\int \frac{d^3k}{2(2\pi)^3}{\rm coth}[ W_{k0}/(2T_i)]
\left[|\dot U_k|^2\Theta(k-2H(0))-{\cal S}^{(1)}\Theta(k-\kappa) 
\right.\nonumber\\&&
\left.+\frac{k^2}{3a^2} (|U_k|^2\Theta(k-2H(0))
-{\cal S}^{(2)}\Theta(k-\kappa)  )
\right]
\nonumber\\
\label{ppluseRk}
\end{eqnarray}
The renormalised energy and pressure should
satisfy the energy conservation equation
\begin{equation}
{\dot  \epsilon}_R({t}) + 3 \, h({t})\, 
(p+\varepsilon)_R({t}) = 0 \; .
\label{energycons}
\end{equation}

For the energy conservation equation to hold the $\Theta$ function for
the subtracted terms should be chosen compatible with the subtraction
in $\langle\psi^2\rangle_R$.  This has been done in Ref. \cite{9610396}
and above.  Refs. \cite{9610396,9709232} differ from each other in their
subtractions for the energy and energy plus
pressure and we find that the expressions
for $\epsilon_R$ and $(p+\epsilon)_R$ in Ref. \cite{9709232} do not
satisfy the energy conservation equation because the lower momentum cutoff
for the different subtracted terms are not chosen appropriately.  We find
that a simple prescription is to adopt the same lower momentum cutoff for
all the subtracted terms in 
$\langle\psi^2\rangle_R$,
 $\epsilon_R$ and $(p+\epsilon)_R$, and not to rescale  $\langle\psi^2\rangle_R$
at $t=0$ to be 0.

An issue that appears to have been overlooked in 
Refs. \cite{9703327,9709232} is that
the WKB solutions obtained above are also valid only for $k^2\gg{\cal M}^2$.
In fact,
one can see that if one guesses a solution for Eq. (\ref{fkeom}) of the form
${\rm exp}[i\int (k^2+{\cal M}^2)^\frac{1}{2} \, d\tau]$ 
then the exponent will approximate 
to a series of the form Eq. (\ref{Rk}) (in the limit of slowly varying
${\cal M}^2$) only if ${\cal M}^2\ll k^2$.
Because of the factor $C(\tau)$ in ${\cal M} $ this condition breaks down 
very quickly for the 
lower momentum modes in the integrals over $k$.
Alternatively, the
validity condition for the WKB solutions is  written as
$\delta^n S_{n+1}\ll \delta^{n-1} S_n$ as 
$\delta\rightarrow0$ \cite{benderorszag}, i.e. succeeding terms in the
series in the exponent of the WKB solution should be smaller than previous
ones.  
Applying this condition to the non-zero terms in the series for $R_k$
we find that it holds only for $k^2\gg{\cal M}^2$.
Similarly, in 
the expansion for $|U_k|^2$ in powers of $1/k$ in Eq. (3.2) of 
Ref. \cite{9610396}, the second term is smaller than the first only if
$k^2/a^2\gg M^2(t)-{\cal R}/6$, and similarly for the second and 
third terms.
Thus the subtractions being carried out do not match the true solution for
increasingly larger numbers of modes as inflation progresses.
(Though the WKB series will not approximate the true solution for $k<\cal M$
this may not interfere with the energy conservation equation as the individual
terms in
the series are obtained consistently 
order by order from the equation of motion.)

Now, for the inflaton field
the dominant 
contribution 
to $\langle \psi^2\rangle_R, \epsilon_R$
and $p_R$ come from low momentum modes.
When a mode becomes unstable, i.e., when its effective
frequency $\omega_{k R}^2$ becomes negative, it grows exponentially fast.
Obviously low momentum modes become unstable earlier and hence start
growing earlier.  Furthermore, after a few e-foldings of inflation
the contribution
of additional modes that become unstable does not add much to the 
contribution of the modes that left in the first few e-foldings.
As discussed above,
for the dominant low momentum modes the WKB solutions
are not valid approximations of their behaviour.  However,
as seen in Fig. 2,
the subtractions carried out in Refs. \cite{9610396,9709232} 
based on the WKB solutions do 
not affect the contribution of these modes much.  This 
may explain why the numerical solutions of Ref. \cite{9709232} 
did not show any conflict with
the energy conservation equation even though the counterterms
were subtracted improperly.
Note that for a field with a positive mass squared, as also considered in
Ref. \cite{9703327}, this explanation will not hold.

{\bf A comparison with the standard prescription}

We now compare the renormalisation schemes adopted in Refs. \cite{9610396,9709232}
based on Ref. \cite{9310319} with the approach outlined in Ref. \cite{vilenkin}.
(See also Ref. \cite{lindebook}.)  The basic difference is that in Ref. \cite{vilenkin}
all modes that have left the horizon during inflation are included in the integral over $k$ in
$\langle \psi^2\rangle$.  This provides an ultraviolet cutoff of $Ha$ for the 
integral.  (In this sub-section we assume $H$ is constant during inflation for
easy comparison with Refs. \cite{vilenkin,lindebook} where this is also 
assumed.)
In the scheme adopted in Refs. \cite{9610396,9709232} all divergent terms in 
the WKB
solution obtained above are subtracted for modes $k\ge\kappa$ from $|U_k|^2$ to obtain
$\langle \psi^2\rangle_R$.  In the former scheme more and more modes contribute to the 
integral in $\langle \psi^2\rangle_R$ as they cross the horizon but this does not seem to
be the goal in the latter scheme.  But it is precisely the contribution of the newer modes
leaving that gives the standard time dependence of $\langle \psi^2\rangle_R$ for a massless or
massive field, as in Ref. \cite{linde}. 

For example, in the case of a massless field for which we know the
exact solution for $U_k$, the prescription of Ref. \cite{vilenkin} gives
\footnote{
As stated in Ref. \cite{vilenkin}, this approach provides a simple but
intuitive way of obtaining $\langle \psi^2\rangle_R$.
A more rigourous derivation with proper regularisation and renormalisation
is given in Ref. \cite{vilenkinford}.}
\begin{eqnarray}
\langle \psi^2\rangle_R&=&\frac{1}{(2\pi)^3}\int_H^{Ha} 4\pi k^2 dk \frac{H^2}{2k^3}
(1+\frac{k^2}{H^2a^2})\cr
&\approx&\frac{H^3t}{4 \pi^2}\, .
\end{eqnarray}
(The second term is ignored as it is suppressed by $a^2$.)
Presuming that the intent
in the scheme of Ref. \cite{9310319} was to subtract the
contribution of modes greater than some renormalisation scale
$\kappa$, and making use of the solution 
obtained in Ref. \cite{vilenkin} to 
correctly subtract the contribution of all modes
$k>\kappa$,
$\langle \psi^2\rangle_R$ will be
\begin{eqnarray}
\langle \psi^2\rangle_R&=&\frac{1}{(2\pi)^3}\int_H^{\kappa} 4 \pi k^2 dk \frac{H^2}{2k^3}
(1+\frac{k^2}{H^2a^2})\cr
&\approx&{H^2\over4\pi^2} ln(\kappa/H)\, .
\end{eqnarray}
(We have chosen the lower momentum cutoff to be $H$ in keeping with
Ref. \cite{vilenkin}.)
One sees that the two prescriptions give very different results which must
be kept in mind while calculating the effect of fluctuations.  

For our field $\psi$ with a negative mass squared
the renormalisation prescription based on Ref. \cite{9310319} implies
\begin{eqnarray}
\langle \psi^2\rangle_R&=&\int_{2H}^\infty
\frac{4\pi k^2 dk}{(2\pi)^3}\frac{|U_k(t)|^2}{2}{\rm coth}[ W_{k0}/(2T_i)]
-
\int_\kappa^\infty
\frac{4\pi k^2 dk}{(2\pi)^3}\frac{|U_K(t)|^2}{2}{\rm coth}[ W_{k0}/(2T_i)]
\label{psi2R1}
\\
&&\cr
&=&\int_{2H}^\kappa
\frac{4\pi k^2 dk}{(2\pi)^3}\frac{|U_k(t)|^2}{2}{\rm coth}[ W_{k0}/(2T_i)]\cr
&&\cr
&&
+
\int_\kappa^{|{\cal M}|}
\frac{4\pi k^2 dk}{(2\pi)^3}\frac{1}{2}(|U_k(t)|^2-|U_K(t)|^2)
{\rm coth}[ W_{k0}/(2T_i)],
\label{psi2R}
\end{eqnarray}   
where $U_K(t)$ are the WKB solutions and $|U_K(t)|^2$ above contains terms of
order $1/k$ and $1/k^3$.  As discussed earlier,
the WKB solutions $U_K(t)$ in the second integrals are not an approximation
for the corresponding $U_k(t)$ for $k\le|{\cal M}|$.  (Eq. (\ref{psi2R}) is written for
times when $\kappa\le |{\cal M}|$, which is all times if $\kappa=m_R$ and after 2 e-foldings
if $\kappa=10m_R$.)  On the other hand,
the renormalisation scheme of Ref. \cite{vilenkin} would
imply
\begin{equation}
\langle \psi^2\rangle_R=\int_{2H}^{Ha}
\frac{4\pi k^2 dk}{(2\pi)^3}\frac{|U_k(t)|^2}{2}{\rm coth}[ W_{k0}/(2T_i)]  
\label{psi2Rlinde}
\end{equation}
(for times when $Ha>2H$, i.e., after 1 e-folding).

Both expressions for $\langle \psi^2\rangle_R$ 
appear different.  
(Since they differ by a finite amount, in principle, both are valid.)  However,
in Fig. 1 we have plotted $\langle \psi^2\rangle_R$ obtained under both renormalisation
schemes and they are similar.  For the case of the inflaton with a negative mass squared
lower momentum modes are growing modes and the contribution of the lowest few modes
$k\le20 m_R$ dominates the integrals in $\langle \psi^2\rangle_R$.  For these modes the WKB
subtractions do not affect their contributions, as seen in Fig. 2.  Therefore 
once $|{\cal M}|$ and $Ha$ become larger than $20 m_R$, which happens
within 3 e-foldings, $\langle \psi^2\rangle_R$ 
obtained under both renormalisation
schemes are similar.

{\bf Some additional comments:}

In Ref. \cite{9610396} the authors cite the possibility 
using the growth
of fluctuations of the inflaton
as a new means of evolving the inflaton in its potential and
providing a graceful exit from the inflationary
phase.  However, 
following the evolution of the fluctuations
rather than the mean field to study the inflationary phase transition
has been studied earlier in Refs. \cite{linde,vilenkin,hawkingmoss2}.

In Sec. V(D) of Ref. \cite{9709232} the authors introduce the notion of
a zero mode assembly consisting of super-horizon sized low momentum modes,
which acts as an effective zero mode after a few e-foldings of inflation
even in the absence of $\phi(t)$.
The authors have also discussed the classicalisation of these low momentum
fluctuations.  We point out that this is similar to ideas discussed in
Sec. IIIA of Ref. \cite{bst} and in Sec. 8.3 of Ref. \cite{lindebook}
in which the early evolution of the inflaton is due to the growth of
quantum fluctuations $\langle \psi^2\rangle_R$
but after some time $t_{cl}$ one can
study inflation by solving the classical equation of a motion for
a purely time dependent field with the initial value 
at $t_{cl}$ taken (in Ref. \cite{lindebook}) to be 
$\langle \psi^2(t_{cl})\rangle_R^{1\over2}$.

\section{Numerical Results}

In this section we present 
our results after numerically solving the large set of
differential equations representing a continuum of $k$-values.
The numerical code is based on fourth order Runge-Kutta for evolution
of the mode functions
and IMSL algorithms for interpolation of
discrete data and integration to obtain $\langle \psi^2\rangle_R$. We have chosen 
$\lambda=10^{-12}$ consistent with the calculation of density 
perturbations in Ref. \cite{9709232} and we set $\xi_R=0$.  
For numerical purposes we follow Ref. \cite{9610396} in introducing the following
dimensionless quantities
\begin{equation}
{\calT} = m_R t \quad ; \quad h= \frac{H}{m_R} \quad ;
\quad q=\frac{k}{m_R} 
\nonumber
\end{equation}
\begin{equation}
g= \frac{\lambda_R}{8\pi^2} 
\quad ; \quad  g\Sigma({\calT}) = \frac{\lambda_R}{2m^2_R}\; \langle 
\psi^2(t)
\rangle_R  \quad ; \quad T_{ii}=T_i/m_R  \; .
\end{equation} 
We treat $H$ as a constant.  Assuming $m_R\sim10^{13}\gev$ we set $h=2$.

In Fig. 1 we present the growth of $g\Sigma$ under the renormalisation
schemes of Ref. \cite{9310319} and Ref. \cite{vilenkin}.
The plot following the renormalisation scheme of Ref. \cite{9310319}
contains the modifications suggested above, namely, 
only modes $k/a_0>2H(0)$ are included in the integrals over $k$,
the theta function for the subtracted
quantities applies to all terms being subtracted and the renormalisation
scale $\kappa$ is taken to be $10 m_R$.
\footnote{
Subsequent figures follow the renormalisation scheme of Ref. \cite{9310319}
with the same modifications.}
It is not possible to distinguish between the two curves and
the quantum fluctuations 
grow dramatically between
${\calT}=60$ and 100 and approach an asymptotic value of 1.

In Fig. 2 we plot $g\Sigma$ with and without the
subtraction of Eq. (\ref{psi2R1}).  The two curves lie on
top of each other.  This implies that the 
effect of the subtraction is
negligible. 

In Fig. 3 we plot $g\Sigma$ for different values of $T_{ii}$.
The initial temperature $T_i$ enters in two places in the 
calculation of the inflaton dynamics in Ref. \cite{9610396}.  Firstly,
it appears in the temperature dependent correction to the 
effective mass for $\psi$ in the initial 
conditions ($r$ is taken to be $T_i/T_c$ in Ref. \cite{9610396})
and secondly, it appears in the expressions for $\langle 
\psi^2\rangle$,
energy, etc. in the factor ${\rm coth}[ W_{k0}/(2T_i)]$.  
To distinguish between these two effects
we choose $T_{ii}=0,200$ and $10^7$ and take $r=0$
for the first case and $r=2$ for the other cases. 
For $T_{ii}=0$,
$g\Sigma$ begins to grow appreciably at ${\calT}\approx 70$ and
achieves 
an asymptotic value of 1.
This agrees well with 
the results in Ref. \cite{9709232}.
For $T_{ii}=10^{7}$,
$g\Sigma$ begins to grow appreciably at ${\calT}\approx 30$. 
This again agrees well with 
the results in Ref. \cite{9610396}.  The corresponding timescale for
$T_{ii}=200$ is 60. 
Comparing the plots for $T_{ii}=200$ and $10^7$ with the same
value of $r$  we see that
it is the 
presence of the coth term representing initial thermal fluctuations of the
$\psi$ field
that plays a substantial role in modifying the duration of inflation.
As a further test we have also varied $r$ in $W_{k0}$ till $r=10$ 
and seen
that there is no significant dependence on $r$.

The inflationary phase transition is completed when the effective
mass for the inflaton modes $-1 + g\Sigma $
achieves a 
value of 0.  On the other hand,
the inflationary era ends when $\ddot a<0$.
Now $\ddot a=-{4\pi\over3}
(\epsilon_R+3 p_R)$ and hence inflation ends when $p_R$ becomes greater
than $-{1\over3}\epsilon_R$.  Since $\dot H/H^2=-{3\over2}
(1+{p_R\over\epsilon_R})$, inflation ends when $\dot H/H^2$ becomes
less than -1.  
Since we do not treat the Hubble constant dynamically we take the 
breakdown of the constant $H$ assumption, when the fluctuations begin
to grow dramatically, to be an indicator of the end of
inflation.  
As the fluctuations grow
the universe will transit from an exponentially inflating universe to
a power law inflating universe and will then enter the reheating era.
Therefore, $h \calT_e$, where $\calT_e$ is the
time when the fluctuations begin to grow dramatically, gives us
a lower
bound on the number of e-foldings of inflation.

Understandably, the existence of frozen-in initial thermal fluctuations 
in $\psi$
decreases the time required for the total fluctuations
$\langle\psi^2\rangle_R$ to reach its final
value.
If one assumes that the inflaton was in thermal equilibrium prior to the onset
of inflation then it is likely it will go out of equilibrium soon after the
onset of inflation and the above shows that it is important to include 
the thermal fluctuations of the initial state while estimating the duration
of inflation.
If the inflaton goes out of thermal equilibrium earlier than the onset
of inflation then the contribution of the thermal fluctuations frozen in at
the higher temperature can be even larger.

\section{Conclusion}

In this paper we have firstly compared the approach of Refs. \cite{9610396,9709232}
with that of earlier works to study the dynamics of the inflaton.  We point out
that for the Hartree and the lowest order large $N$ approximations
employed in Refs. \cite{9610396,9709232} with
the CTP formalism
the equations of motion are akin to those obtained in standard field theory.
We have reobtained the large momentum mode solutions for the inflaton field
using the WKB method and pointed out some differences with earlier results.
Because of these differences and the very involved nature of these calculations
we have included many explicit details of the calculations.  We have also included
temperature corrections in the initial effective mass of the inflaton field
and pointed out that the assertion in Refs. \cite{9703327,9709232} that
their choice of initial conditions in conformal spacetime 
gives
counterterms that are initial conditions independent
is only true if there are no thermal corrections to the initial inflaton mass.
We have discussed different aspects 
of the renormalisation prescription adopted 
in Refs. \cite{9610396,9709232}.
In particular, we have pointed out that the asymptotic expressions used
in Refs. \cite{9610396,9709232} for the large momentum modes during 
renormalisation are valid for large $k/a$ and not large $k$.
We have also compared the renormalisation prescription
of Refs. \cite{9610396,9709232} with that of earlier works and pointed 
out their differences.
Despite these differences the results of these different approaches are the same
because for an inflaton with a negative mass squared it is the low momentum modes,
and not the high momentum modes regularised by renormalisation, 
that give the dominant 
contribution.

\acknowledgements
We would like to thank D. R. Kulkarni for
advice on the numerical codes.

\newpage
\centerline{\bf Figure Captions:}

{\bf Fig. 1:} Renormalised fluctuations of the inflaton field, 
$g\Sigma$, are 
shown
as a function of normalised time ${\bar t}$ for the two
renormalisation schemes discussed in the text (with constant $H$). 
$r=2$, $h=H/m_R=2$  and the normalised initial temperature
$T_{ii}=T_i/m_R=200$. 
The solid line corresponds to the (modified) renormalisation
scheme of Ref. \cite{9310319} while the points correspond to the
renormalisation scheme of Ref. \cite{vilenkin}.

\vspace{2mm}

{\bf Fig. 2:} Fluctuations of the inflaton field, with and without the
WKB subtractions (solid line and circles respectively),
are 
shown
as a function of normalised time ${\bar t}$.
All parameters are as in Fig. 1.

\vspace{2mm}

{\bf Fig. 3:} 
$g\Sigma$ vs. $\calT$ for
$T_{ii}=0,200$ and $10^7$.
$r=2$ for $T_{ii}=200$ and $10^7$
and $r=0$ for $T_{ii}=0$.  $h=2$ for all plots.

\newpage
\psfig{figure=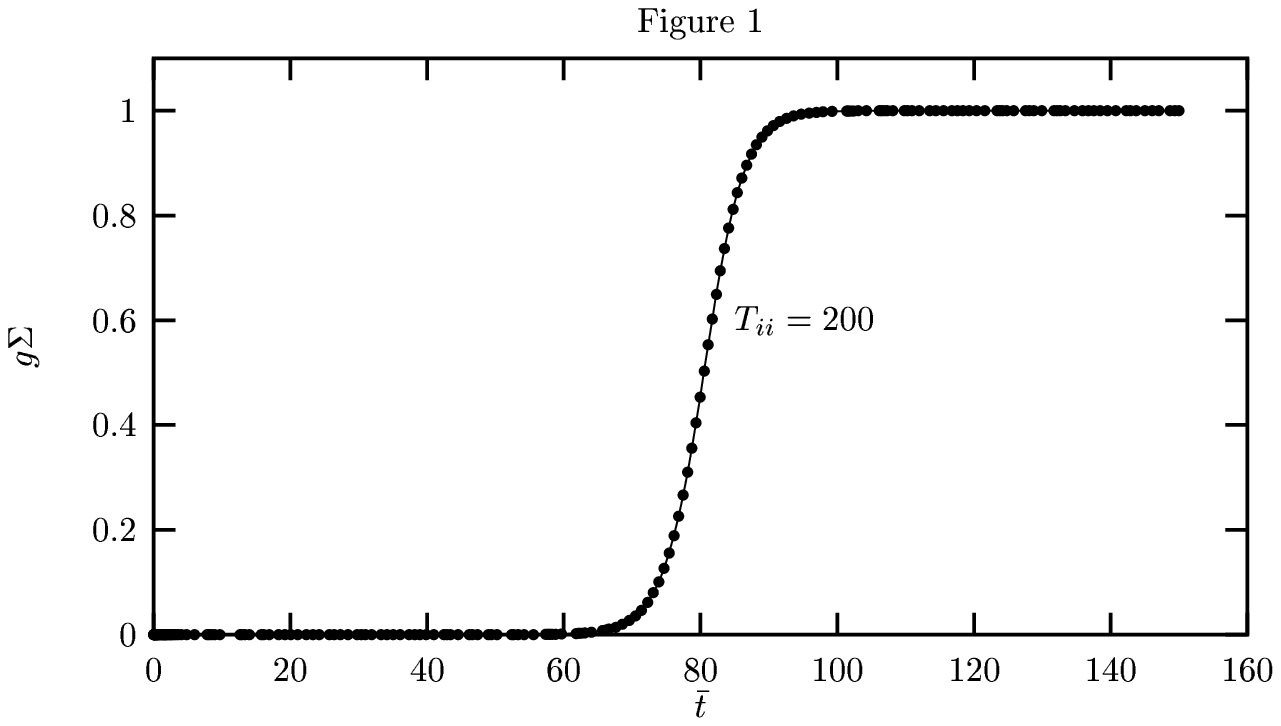}
\newpage
\psfig{figure=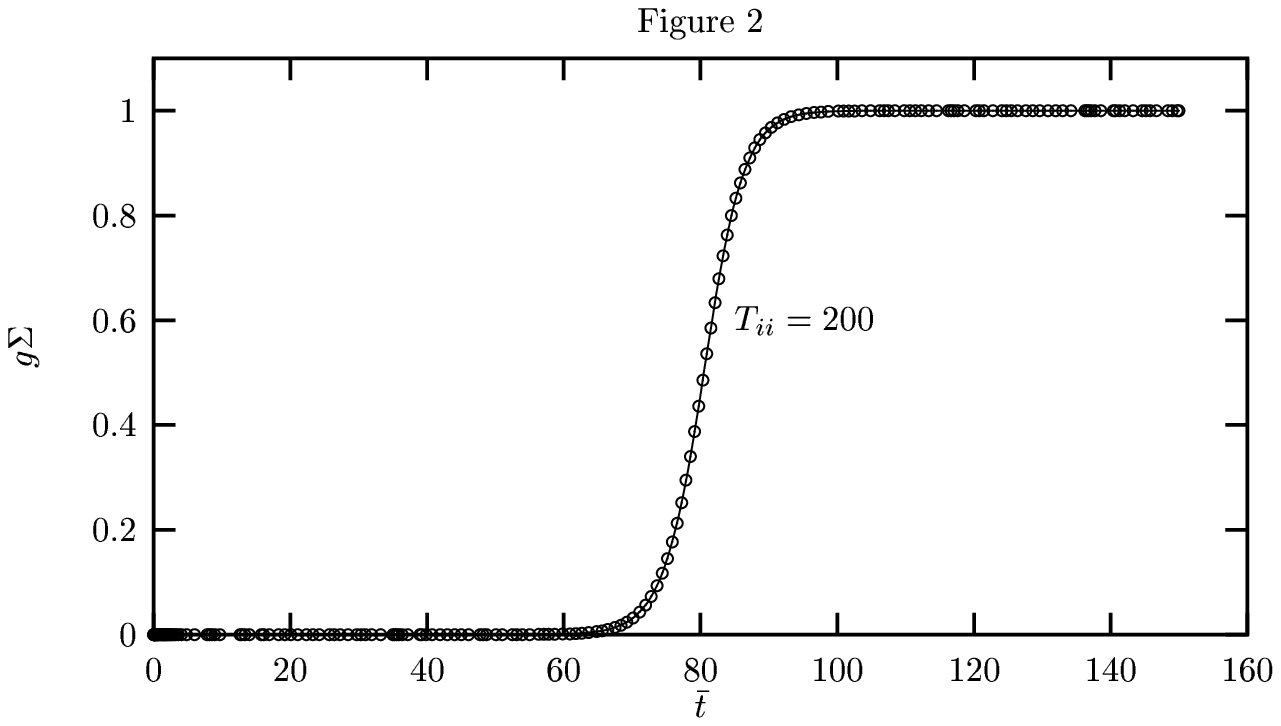}
\newpage
\psfig{figure=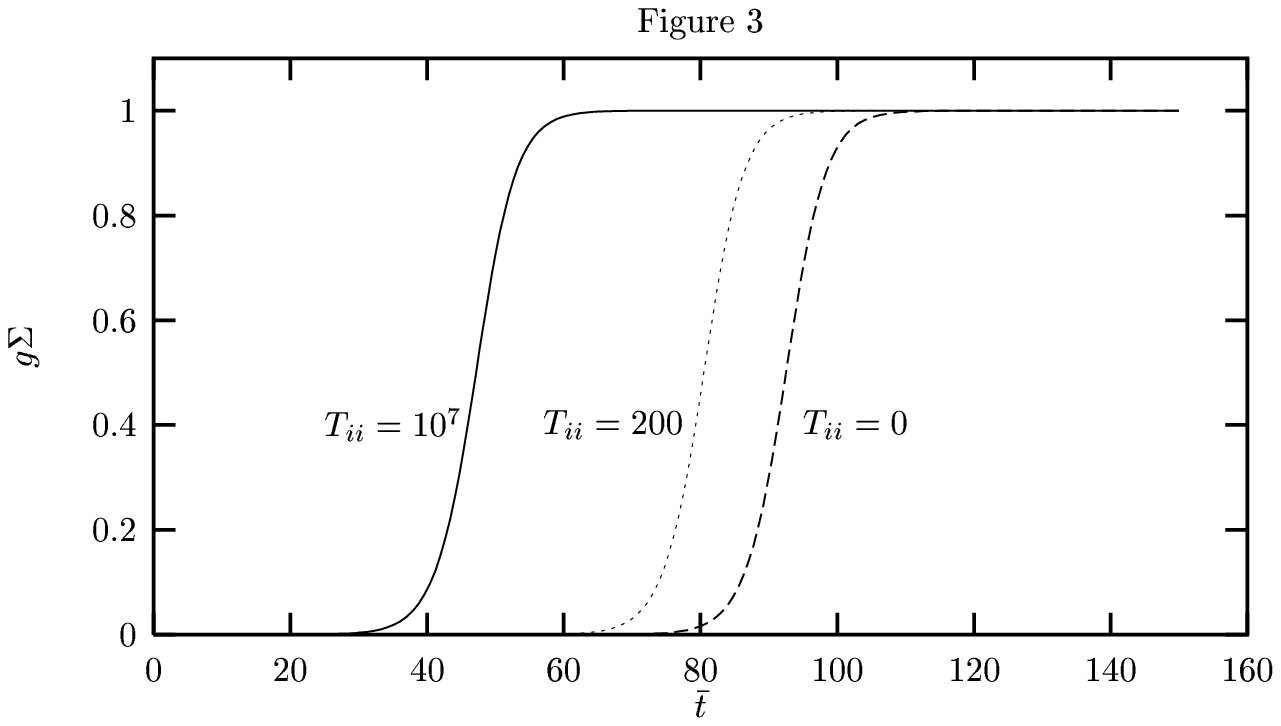}

\end{document}